\documentstyle[12pt]{article}

\title{
\begin{flushright}
{\normalsize Yaroslavl State University\\
             Preprint YARU-HE-95/04\\
             hep-ph/9511366} \\[5mm]
\end{flushright}
Unitary Limit on $K^0_L \to \mu^+ \mu^-$ and \\
       the Top Quark Mass \\[5mm]}

\author{Gvozdev A.A., Mikheev N.V. and Vassilevskaya L.A.\\
        {\it Yaroslavl State University, Yaroslavl 150000, Russia} \\[25mm]}

\date{}

\begin{document}

\maketitle

\large

\begin{abstract}
A brief overview of the recent measurements of the branching ratio
of the rare $K_L^0 \to \mu^+ \mu^-$ decay in the context of their
agreement with the Standard Model (SM) is given. It is shown that
KEK results well correlate with the SM and B-physics, whereas the
BNL results are in conflict with the SM with the heavy top quark.
\end{abstract}

\newpage

For quite a long time the rare electroweak decays
$K^0_L \to \mu^+ \mu^-$ and $K^0_L \to \gamma \gamma$ have been
the subject of intensive in\-ves\-ti\-ga\-tions.
An observation of the heavy top quark with the mass
$\sim 190 \, GeV$~\cite{CDF,DO}
and a further precision of the measurement of $V_{u b}$ and $V_{c b}$
CKM matrix elements improved by ARGUS and CLEO~\cite{Ali}
make a curious situation about
the rare electroweak decay $K^0_L \to \mu^+ \mu^-$.
It is known that the connection between the absorptive part of the
$K^0_L \to \mu^+ \mu^-$ decay width
and $K^0_L \to \gamma \gamma$ decay width
gives us the down limit of the probability of the
$K^0_L \to \mu^+ \mu^-$ decay:

\begin{equation}
Br_{abs} ( K^0_L \to \mu^+ \mu^- ) \simeq 1.2 \cdot 10^{-5} \,
Br ( K^0_L \to \gamma \gamma ) = (6.8 \pm 0.3) \cdot 10^{-9} .
\label{eq:ur}
\end{equation}

\noindent Here we used the experimental value of
$Br ( K^0_L \to \gamma \gamma ) = (5.73 \pm 0.27) \cdot 10^{-4}$~\cite{PDG}.
This minimal value allowed by theory is known  as the
unitarity limit.

In our papers~\cite{GMV} we have investigated the processes
$K^0_L \to \mu^+ \mu^-$ and $K^0_L \to \gamma \gamma$
within the quark model approach to obtain the estimation
of the top quark mass.
In this approach the $K^0_L \to \mu^+ \mu^-$ amplitude is the sum of
one-loop ($1L$) and two-loop ($2L$) contributions. The first one
(through $W$ and $Z$) due to short distances $\sim 1/m_W$ where the top
quark contribution dominates. As for the $2L$ contribution with two
photons in the intermediate state, both intermediate ($\sim 1/m_c$)
and rather long
($\ge 1/m_K$) distances are essential in it.

We pointed out the principal importance of the correct relative sign
of the $1L$ and $2L$ contributions in the amplitude.
Let us note that in the terms of the bare quarks the total decay
amplitude contains these contributions with opposite signs~\cite{VSh}.
However we emphasize that it is necessary to account
the QCD corrections to the
effective four-quark vertex to obtain a realistic result for $2L$
contribution. To that end we used the renormgroup method by Vainstein,
Zakharov and Shifman~\cite{VZSh}
for the mass scale $\mu$ down to
the typical hadronic scale $\mu_0 \simeq 2 \Lambda$
($\alpha_{st}(\mu_0) = 1$).
We have developed also a fenomenological method of the estimation of the
QCD corrections in the region $\mu_0 \le \mu \le m_K$~\cite{GMV}.
To test the reliability of our method we
estimated the ratio $\Gamma(K_L^0 \to e^+ e^- \gamma) /
\Gamma(K_L^0 \to \mu^+ \mu^- \gamma)$~\cite{G3}
and showed that our result
was in agreement with one obtained within
the phenomenological pole model~\cite{B}.
Certainly we do not pretend to obtain an integral accuracy better than
$30 \%$ in the description of the contributions of relatively
long distances ($r \le 1/ \mu_0$). However, the sign between the $1L$ and
the real part of the $2L$ contributions is fixed sufficiently reliable by
this way. Our main result is that the real part of the $2L$ contribution
changes the sign if the QCD corrections are taken into account.
The change of the sign is connected with the behaviour of the integral
over the $u$-quark loop scale. This integral involves multiplicatively
the QCD formfactor of an effective four-quark
 $(V-A)$ vertex which becomes
sufficiently large (more than unit in modulus) and negative
number in the region $2 \Lambda \le \mu \le m_K$~\cite{VZSh}.

The expression for the total $K_L^0 \to \mu^+ \mu^-$
amplitude obtained by this way
has the form~\cite{GMV}:

\begin{eqnarray}
{\cal M} ( K_L^0 \to \mu^+ \mu^- ) & \simeq & -  10^{-3} \, {\cal N} \,
\big \lbrace + [ ( 5.6 \pm 2.0 ) - i (44.7 \pm 0.9) ] \nonumber \\
&& + 2 + 10^3 \; \frac{F(m_t^2 / m_W^2)}{\sin^2 \theta_W} \;
\frac{\Re (V^*_{t d} V_{t s})}{\Re (V^*_{c d} V_{c s})} \big \rbrace ,
\label{eq:ampl} \\
F(x) & = & \frac{x}{4} \, \left [ \frac{4 - x}{1 - x} +
\frac{3 x \ln x}{(1 - x)^2} \right ] ,  \nonumber
\end{eqnarray}

\noindent where ${\cal N} = (\alpha / 4 \pi) \, G_F \, F_K \, m_\mu
\sin 2 \theta_C (\bar \mu \gamma_5 \mu)$,
$F_K$ is the formfactor of the $K$-meson, $m_\mu$ is
the muon mass, $\theta_C$ and $\theta_W$ are the Cabibbo and Weinberg
angles, respectively;
$F(x)$ is the well-known function~\cite{MP},
The first term in the curly braces describes
the $2L$ contribution.
We pretend only on the calculation of the real part of the $2L$ contribution,
and take the imagine part from the unitarity relation~(\ref{eq:ur}).
The second and third terms of the amplitude~(\ref{eq:ampl}) describe the $c$-
and $t$-quark contributions, respectively.

It should be noted that our expression~(\ref{eq:ampl}) for the total
decay amplitude is in contradiction with the calculation of Ko~\cite{Ko}
in which short and long distance contributions have opposite signs.
The method developed by Ko~\cite{Ko}
has, in our opinion, disadvantage.
Namely, in these papers the dependence of the meson vertex
formfactors (for example, $\pi V V$) on the meson loop scale
was neglected.

To obtain the restriction on the $K_L^0 \to \mu^+ \mu^-$ decay width,
we used the resent experimental data on the top quark mass
of the CDF Collab.~\cite{CDF}

\begin{displaymath}
m_t = 176 \pm 8 \mbox{(stat.)} \pm 10 \mbox{(sys.)} \, GeV/c^2
\end{displaymath}

\noindent and on the parameters of CKM matrix
in the Wolfenstain representation~\cite{Ali}

\begin{eqnarray}
&& \lambda = \sin \theta_C \simeq 0.22 , \qquad A = 0.80 \pm 0.12,
\nonumber \\
&& \sqrt{\rho^2 + \eta^2} = 0.36 \pm 0.14 \qquad \mbox{which gives} \qquad
(1 - \rho) \ge 0.64 \pm 0.14.
\nonumber
\end{eqnarray}

On Fig.1 we present the experimental data of the mea\-su\-re\-ment
of $Br(K_L^0 \to \mu^+ \mu^-)$
of BNL E791 Collab.~\cite{BNL}
and KEK E137 Collab.~\cite{KEK}.
%
%
\begin{figure}[tb]
\unitlength=1mm
\special{em:linewidth 0.4pt}
\linethickness{0.4pt}
\begin{picture}(150.00,70.00)(-30.00,00.00)

\put(55.00,5.00){\makebox(0,0)[cc]{\large Figure 1.}}

\put(0.00,20.00){\line(0,1){45.00}}
\put(0.00,20.00){\line(1,0){115.00}}

\multiput(0.00,20.00)(10.00,0.00){11}{\line(0,-1){2.00}}
\multiput(0.00,20.00)(0.00,5.00){9}{\line(-1,0){2.00}}

\put(0.00,65.00){\line(-1,-4){1.00}}
\put(0.00,65.00){\line(1,-4){1.00}}
\put(115.00,20.00){\line(-4,1){4.00}}
\put(115.00,20.00){\line(-4,-1){4.00}}

\put(-7.00,20.00){\makebox(0,0)[lc]{5}}
\put(-7.00,30.00){\makebox(0,0)[lc]{6}}
\put(-7.00,40.00){\makebox(0,0)[lc]{7}}
\put(-7.00,50.00){\makebox(0,0)[lc]{8}}
\put(-7.00,60.00){\makebox(0,0)[lc]{9}}
\put(3.00,65.00){\makebox(0,0)[lc]{$Br(K_L^0 \to \mu^+ \mu^-) \cdot 10^9$}}

\put(00.00,13.00){\makebox(0,0)[cb]{86}}
\put(20.00,13.00){\makebox(0,0)[cb]{88}}
\put(40.00,13.00){\makebox(0,0)[cb]{90}}
\put(60.00,13.00){\makebox(0,0)[cb]{92}}
\put(80.00,13.00){\makebox(0,0)[cb]{94}}
\put(100.00,13.00){\makebox(0,0)[cb]{96}}
\put(110.00,13.00){\makebox(0,0)[lb]{\large year}}

\put(0.00,35.00){\line(1,0){108.00}}
\put(0.00,38.00){\line(1,0){108.00}}
\put(0.00,41.00){\line(1,0){108.00}}
\multiput(0.00,35.00)(6.00,0.00){1}{\line(1,1){6.00}}
\multiput(0.00,41.00)(6.00,0.00){1}{\line(1,-1){6.00}}
\multiput(36.00,35.00)(6.00,0.00){12}{\line(1,1){6.00}}
\multiput(36.00,41.00)(6.00,0.00){12}{\line(1,-1){6.00}}
\put(9.00,38.00){\makebox(0,0)[lc]{Unitary limit}}

\put(0.00,49.00){\line(1,0){108.00}}
\multiput(0.00,49.00)(10.00,0.00){11}{\line(1,1){10.00}}
\put(55.00,51.00){\makebox(0,0)[lb]{SM with heavy $t$-quark}}

\put(94.00,64.00){\rule{2.00\unitlength}{2.00\unitlength}}
\put(98.00,65.00){\makebox(0,0)[lc]{BNL}}
\put(95.00,60.00){\circle*{2.00}}
\put(98.00,60.00){\makebox(0,0)[lc]{KEK}}


\put(25.00,46.50){\line(0,1){14.00}}
\put(24.00,46.50){\line(1,0){2.00}}
\put(24.00,60.50){\line(1,0){2.00}}
\put(25.00,53.50){\circle*{2.00}}

\put(40.00,42.50){\line(0,1){14.00}}
\put(39.00,42.50){\line(1,0){2.00}}
\put(39.00,56.50){\line(1,0){2.00}}
\put(40.00,49.50){\circle*{2.00}}

\put(30.00,34.00){\line(0,-1){16.00}}
\put(29.00,34.00){\line(1,0){2.00}}
\put(29.00,18.00){\line(1,0){2.00}}
\put(29.00,25.00){\rule{2.00\unitlength}{2.00\unitlength}}

\put(45.00,39.00){\line(0,1){8.00}}
\put(44.00,39.00){\line(1,0){2.00}}
\put(44.00,47.00){\line(1,0){2.00}}
\put(44.00,42.00){\rule{2.00\unitlength}{2.00\unitlength}}

\put(75.00,36.00){\line(0,1){6.00}}
\put(74.00,36.00){\line(1,0){2.00}}
\put(74.00,42.00){\line(1,0){2.00}}
\put(74.00,38.00){\rule{2.00\unitlength}{2.00\unitlength}}

\put(95.00,35.50){\line(0,1){5.00}}
\put(94.00,35.50){\line(1,0){2.00}}
\put(94.00,40.50){\line(1,0){2.00}}
\put(94.00,37.00){\rule{2.00\unitlength}{2.00\unitlength}}

%
\end{picture}
\end{figure}
%
%
As illustrated in Fig.1, the KEK results well correlate
with the SM and $B$-physics, whereas the BNL results
are in conflict with the SM with the heavy $t$-quark.
If the BNL result is verified
by new series of more precise measurements it may be a signal
of a new physics beyond the SM. For example, the real part of the total
amplitude~(\ref{eq:ampl}) can contain a contribution
of the relatively light leptoquark~\cite{KM}
which can cancel sufficiently the contribution of the heavy top quark.

The research described in this publication was made possible in part by
Grant No. RO3300 from the International Science Foundation and Russian
Government.
The work of Mikheev N.V. was supported by a Grant N d104
by International Soros Science Education Program.
The work of Vassilevskaya L.A. was supported by a fellowship
of INTAS Grant 93-2492 and is carried out within the research
program of International Center for Fundamental Physics in Moscow.

\end{document}